\documentclass[12pt]{article}%
\usepackage{amsmath}
\usepackage{amsfonts}
\usepackage{amssymb}
\usepackage{graphicx}%
\providecommand{\U}[1]{\protect\rule{.1in}{.1in}}
\newcommand{\bfr}{\begin{flushright}}
\newcommand{\efr}{\end{flushright}}
\newcommand{\bc}{\begin{center}}
\newcommand{\ec}{\end{center}}
\newcommand{\ben}{\begin{enumerate}}
\newcommand{\een}{\end{enumerate}}

\newcommand{\be}{\begin{equation}}
\newcommand{\ee}{\end{equation}}
\newcommand{\ba}{\begin{array}}
\newcommand{\ea}{\end{array}}
\def\6{\partial}
\begin{document}

\title{\textbf{Noncommutative fluid dynamics} \\\textbf{in the Snyder space-time} }
\author{M. C. B. Abdalla$^{a}$\thanks{email: mabdalla@ift.unesp.br},\thinspace\ L.
Holender$^{b}$\thanks{email: holender@ufrrj.br},\thinspace\ M. A. Santos$^{c}%
$\thanks{email: masantos@cce.ufes.br}\thinspace\ and I. V. Vancea$^{b}%
$\thanks{email: ionvancea@ufrrj.br}\\$^{a}$\emph{{\small Instituto de F{\'{\i}}sica Te\'{o}rica, UNESP -
Universidade Estadual Paulista,}}\\\emph{{\small Rua Dr. Bento Teobaldo Ferraz 271, Bloco 2, Barra-Funda,}}\\\emph{{\small Caixa Postal 70532-2, 01156-970, S\~ao Paulo - SP, Brasil}}\\$^{b}$\emph{{\small Grupo de F{\'{\i}}sica Te\'{o}rica e Matem\'{a}tica
F\'{\i}sica, Departamento de F\'{\i}sica,}}\\\emph{{\small Universidade Federal Rural do Rio de Janeiro (UFRRJ),}}\\\emph{{\small Cx. Postal 23851, BR 465 Km 7, 23890-000 Serop\'{e}dica - RJ,
Brasil }}\\$^{c}$\emph{{\small Departamento de F\'{\i}sica e Qu\'{\i}mica,}}\\\emph{{\small Universidade Federal do Esp\'{\i}rito Santo (UFES),}}\\\emph{{\small Avenida Fernando Ferarri S/N - Goiabeiras, 29060-900 Vit\'{o}ria
- ES, Brasil}}}
\date{13 July 2012}
\maketitle

\thispagestyle{empty}
%\pagestyle{empty}

%\begin{center}
%\rule{15cm}{0.01cm}
%\end{center}

\abstract{In this paper, we construct for the first time the noncommutative fluid with the deformed Poincar\'{e}
invariance. To this end, the realization formalism of the noncommutative spaces is employed and the results are
particularized to the Snyder space. The noncommutative fluid generalizes the fluid model in the action functional
formulation to the noncommutative space. The fluid equations of motion and the conserved energy-momentum
tensor are obtained.}

%\begin{center}
%\rule{15cm}{0.01cm}
%\end{center}
\vfill

%\begin{flushright}
%{\footnotesize \textbf{DEFIS-ICE-UFRRJ/2011 \hspace{0.5cm} TH-PHYS/02} }
%\end{flushright}

%-------------------------------------------------------------------------------
\newpage\pagestyle{plain} \pagenumbering{arabic}

\section{Introduction}

Recent studies have shown that physical systems from a variety of fields have
physical properties that can be cast simultaneously in terms of concepts from
two distinct areas: the noncommutative gauge theories and the fluid mechanics
\cite{Bahcall:1991an,Susskind:2001fb,ElRhalami:2001xf,Barbon:2001dw,Barbon:2007zz,Polychronakos:2007df}%
. This leads to the natural question whether there is a well defined
noncommutative fluid theory. Probably the best known example of a system in
which the two fields are closely related is the quantum Hall liquid whose
granular structure can be described in terms of noncommutative gauge fields.
In particular, the quantum Hall effect for fraction $1/n$, the abelian
noncommutative Chen-Simons theory at level $n$ and the Laughlin theory at
$1/n$ are related by a mapping among noncommutative spaces.
\cite{Bahcall:1991an,Susskind:2001fb}. The comparison of the transformations
of the fluid phase space and of symmetries of the noncommutative field
theories suggests that there should be a deeper analogy between the volume
preserving diffeomorphisms in the commutative phase space and the symplectic
preserving diffeomorphism in the noncommutative space that might lead to the
noncommutative analogue of the Bernoulli equation
\cite{Jackiw:2002pn,Jackiw:2002tw,Jackiw:2003dw,Jackiw:2004nm}. More arguments
in favor of the noncommutative fluids can be found in \cite{Alavi:2006sr}
where it was shown that the lowest Landau levels of the charged particles are
related to the noncommutativite curvilinear coordinate operators and in
\cite{DeFelice:2009bx} where the linear cosmological perturbations of a
quantum fluid were shown to exhibit noncommutative properties. In
\cite{Marcial:2010zza}, a generalization of the symplectic structure of the
irrotational and rotational non-relativistic fluids to the noncommutative
space was proposed.

The generalization of the fluid equations to define noncommutative fluids it
is not obvious since many extensive long range degrees of freedom of the
commutative systems do not have a simple interpretation in terms of quantities
defined on the noncommutative spaces. Therefore, finding the noncommutative
correspondents of the statistical mechanics or thermodynamical concepts is a
non-trivial open problem which has not been fully undertaken in the literature
(see for some tentative approaches \cite{Alavi:2007dr,Huang:2009yx}). However,
there are canonical formulations of the ideal fluids in terms of the
Lagrangian functionals over the set of fluid potentials \cite{Jackiw:2000mm}
that can be generalized to noncommutative functionals. This procedure needs to
be supplemented by a correspondence principle needed to fix the constraints to
be imposed on the noncommutative fluid fields in order to obtain the known
fluid equations in the commutative limit. By pursuing this line of reasoning,
some of us have proposed a noncommutative fluid action in terms of the Moyal
deformed algebra of functions over the Minkowski space-time $\mathcal{M}$
\cite{Holender:2011px} that generalizes the commutative relativistic ideal
fluid in the K\"{a}hler parametrization \cite{Nyawelo:2003bv} in which the
fluid is parametrized in terms of one real $\theta(x)$ and two complex
potentials $z(x)$ and $\bar{z}(x)$, respectively. As shown in
\cite{Jackiw:2000mm}, the description of the fluid degrees of freedom in terms
of fluid potentials allows one to lift the obstruction to inverting the
symplectic form in the canonical phase space of the fluid variables. (For
other applications of the K\"{a}hler parametrization of the fluid potentials
see
\cite{Nyawelo:2003bv,Baleanu:2004sc,Jarvis:2005hp,Nyawelo:2003bw,Grassi:2011wt,Holender:2008qj,Holender:2012uq}%
.)

The action functional from \cite{Holender:2011px} describes the noncommutative
fluid model on \emph{canonical} \emph{noncommutative} \emph{space-times}
\cite{AmelinoCamelia:2001fd}, i. e. spaces with the coordinate algebra
characterized by a constant antisymmetric matrix $\theta_{\mu\nu}$. However,
the canonical coordinate algebra and the Lorentz algebra are inconsistent with
each other. Therefore, a \emph{Lie-algebra noncommutative space} structure is
needed \cite{AmelinoCamelia:2001fd} in order to properly generalize the
relativistic fluid to a noncommutative Lorentz covariant model. One
interesting alternative is the Snyder space $\mathcal{S}$ \cite{Snyder:1946qz}
in which the noncommutative coordinates are interpreted as the Lie generators
of $so(1,4)/so(1,3)$. The algebra of functions over the Snyder $\mathcal{F}%
(\mathcal{S})$ space can be endowed with the star-product and the co-product
constructed recently in \cite{Girelli:2009ii,Girelli:2010wi,Banerjee:2006} and is isomorphic
to the deformed algebra over the Minkowski space-time $\left(  C^{\infty
}(\mathcal{M}),\star\right)  $. However, the formulation of the field theory
in the Snyder space is not trivial, since the star product is nonassociative
and the momenta associated to the coordinates do not form a Lie group. A
particularly important problem for these systems is to define and calculate
the relevant physical quantities such as the energy and the linear momenta.

In this paper, we are going to construct the noncommutative fluid in the
Snyder space-time by generalizing the Lagrangian functional approach from
\cite{Holender:2011px}. To this end, we found convenient to formulate the
geometry of $\mathcal{S}$ in the\emph{ realization formalism} developed in
\cite{Jonke:2001xk,Meljanac:2006ui,KresicJuric:2007nh,Meljanac:2007xb,Govindarajan:2008qa,Battisti:2008xy,Govindarajan:2009wt,Meljanac:2009fy}
(see for similar ideas \cite{Lukierski:1993wx,Ghosh:2007ai}) that has been
used recently in \cite{Battisti:2010sr} in an attempt to formulate the scalar
field theory. The realization method has at least two nice features: it allows
one to circumvent the problems related to the nonassociativity in the
interacting field theories and it represents an unified framework for handling
simultaneously different types of noncommutative spaces such as the Snyder,
the Maggiore and the Weyl spaces, respectively, \cite{Battisti:2010sr}. Also,
it can be used to interpolate between the $k$-deformed Minkowski and the
Snyder space-times \cite{Meljanac:2011mt}. In the realization formalism, the
coordinates belong to the noncommutative space in which the algebra of the
coordinate operators closes over the generators of the Lorentz symmetry. The
corresponding momenta are defined as being the duals to the coordinates and
they belong to a coset space. In general, the algebra of coordinates does not
fix the commutation relations either among the momenta or among the
coordinates and momenta. In order to obtain a noncommutative fluid with the
largest symmetry group, we require that the symmetries of the noncommutative
space-time be described by the undeformed Poincar\'{e} algebra. Also, we
require that the commutative limit of the noncommutative fluid be the
relativistic ideal fluid in the Clebsh parametrization in which the fluid
potentials are given in terms of three real fields $\theta(x)$, $\alpha(x)$
and $\beta(x)$, respectively, and they parametrize the velocity of the fluid
elements as $v_{\mu}=\partial_{\mu}\theta+\alpha\partial_{\mu}\beta$
\cite{Jackiw:2000mm}. The present construction can be easily applied to the
relativistic fluid in the K\"{a}hler parametrization.

The paper is organized as follows. In Section 2 we review the geometry of the
Snyder space-time in the realization formalism and establish our notations. In
Section 3 we construct the noncommutative Lagrangian that generalizes the
relativistic ideal fluid in the Clebsch parametrization. We discuss the
transformation of the action under the symmetries of the noncommutative space.
The energy-momentum tensor is defined from the variation of the action under
the deformed translations and we show that it satisfies a conservation
equation. The last section is devoted to conclusions.

\section{Geometry of the Snyder space-time}

In this section we are going to review the geometry of the Snyder space-time
in the framework of the realization formalism following
\cite{Meljanac:2007xb,Battisti:2010sr}. The Snyder space-time is a lattice
space characterized by a length scale $l_{s}$ and compatible with the Lorentz
symmetry. These two properties are obtained by associating noncommutative
position operators $\tilde{x}_{\mu}$ to the sites of the lattice. The algebra
of $\tilde{x}_{\mu}$'s is closed over the generators of $so(1,3)$. The Snyder
algebra was originally obtained by descending from five dimensions and can be
interpreted as a deformed algebra of the $so(1,3)$ with the deformation
parameter $s=l_{s}^{2}$ \cite{Snyder:1946qz}.

Let us start with the deformed algebra generated by the operators $\left\{
\tilde{x}_{\mu},p_{\mu},M_{\mu\nu}\right\}  $ that satisfy the following
commutation relations%
\begin{align}
\left[  \tilde{x}_{\mu},\tilde{x}_{\nu}\right]   &  =sM_{\mu\nu}%
,\label{alg-Snyder-x}\\
\left[  p_{\mu},p_{\nu}\right]   &  =0,\label{alg-Snyder-p}\\
\left[  M_{\mu\nu},M_{\rho\sigma}\right]   &  =\eta_{\nu\rho}M_{\mu\sigma
}-\eta_{\mu\rho}M_{\nu\sigma}+\eta_{\mu\sigma}M_{\nu\rho}-\eta_{\nu\sigma
}M_{\mu\rho},\label{alg-Snyder-M}\\
\left[  M_{\mu\nu},\tilde{x}_{\rho}\right]   &  =\eta_{\nu\rho}\tilde{x}_{\mu
}-\eta_{\mu\rho}\tilde{x}_{\nu},\label{alg-Snyder-Mx}\\
\left[  M_{\mu\nu},p_{\rho}\right]   &  =\eta_{\nu\rho}p_{\mu}-\eta_{\mu\rho
}p_{\nu}, \label{alg-Snyder-Mp}%
\end{align}
where $\mu,\nu=\overline{0,3}$ and the deformation parameter is $s>0$. The
generators $M_{\mu\nu}$ satisfy the commutation relations of the Lorentz group
and can be written in terms of the commutative coordinates of the underlying
Minkowski space-time $\mathcal{M}$ in the usual way $M_{\mu\nu}=i(x_{\mu
}p_{\nu}-x_{\nu}p_{\mu})$.\ Thus, the Snyder algebra defined by the
commutators\ (\ref{alg-Snyder-x})-(\ref{alg-Snyder-Mp})\ can be interpreted as
a deformation of the commutative Poicar\'{e} algebra of $\mathcal{M}$. In
fact, the relation (\ref{alg-Snyder-x}) shows that\ the noncommutative
coordinates $\tilde{x}_{\mu}$\ are functions of the commutative phase space
variables $x_{\mu}$ and $p_{\mu}$. However, the Snyder algebra leaves the
functions $\tilde{x}_{\mu}(x,p)$ and the commutators $[\tilde{x}_{\mu
}(x,p),p_{\nu}]$ undetermined\footnote{It has been shown in
\cite{Battisti:2008xy} that there are infinitely many commutation relations
among $\tilde{x}_{\mu}$ and $p_{\nu}$ that are compatible with the Snyder
algebra.}. The \emph{realizations} of the noncommutative Snyder geometry are
defined by the simplest choice possible for the coordinate operators
$\tilde{x}_{\mu}(x,p)$ as momentum dependent rescalings of the coordinates
$x_{\mu}$%
\begin{equation}
\tilde{x}_{\mu}(x,p)=\Phi_{\mu\nu}(s;p)x_{\nu}. \label{representations-1}%
\end{equation}
The smooth functions $\Phi_{\mu\nu}(s;p)$ can be reduced to a set of two
dependent functions $\varphi_{1}$and $\varphi_{2}$%
\begin{align}
\tilde{x}_{\mu}(x,p)  &  =x_{\mu}\varphi_{1}(A)+s\left\langle xp\right\rangle
p_{\mu}\varphi_{2}(A),\label{rep-1}\\
\varphi_{2}(A)  &  =\left[  1+2\frac{d\varphi_{1}(A)}{dA}\right]  \left[
\varphi_{1}(A)-2A\frac{d\varphi_{1}(A)}{dA}\right]  ^{-1}, \label{rep-2}%
\end{align}
and $A=s\eta^{\mu\nu}p_{\mu}p_{\nu}$. The commutative scalar product is
denoted by $\left\langle ab\right\rangle =\eta^{\mu\nu}a_{\mu}b_{\nu}$. The
realizations defined by the relations (\ref{representations-1})-(\ref{rep-2})
show that the Snyder geometry defined by the algebra (\ref{alg-Snyder-x}%
)-(\ref{alg-Snyder-Mp}) can be viewed as a non-canonical deformation of the
commutative phase space. Different realizations can be obtained by choosing
different functions $\varphi_{1}(A)$. For example, the Weyl, the Maggiore and
the Snyder noncommutative spacetimes can be obtaining by choosing $\varphi
_{1}(A)=\sqrt{A}\cot(A)$, $\varphi_{1}(A)=\sqrt{1-sp^{2}}$ and $\varphi
_{1}(A)=1$, respectively \cite{Battisti:2010sr}. The physical momenta depend
on the specific realization since%
\begin{equation}
\tilde{p}_{\mu}=f(A)p_{\mu},\qquad f(A)=\left[  \varphi_{1}(A)^{2}+A\right]
^{-\frac{1}{2}}. \label{momenta}%
\end{equation}
An important property is that from the point of view of the realizations, the
algebras generated by $\left\{  \tilde{x}_{\mu},p_{\mu},M_{\mu\nu}\right\}  $
are deformed Heisenberg algebras%
\begin{equation}
\left[  \tilde{x}_{\mu},p_{\nu}\right]  =i(\eta_{\mu\nu}\varphi_{1}%
(A)+sp_{\mu}p_{\nu}\varphi_{2}(A)). \label{alg-Snyder-xp}%
\end{equation}
The symmetries of the Snyder space-time are described by the algebra of the
Lorentz generators and the co-algebra of the translation generators acting on
the noncommutative coordinates $\tilde{x}_{\mu}$ according to the relations
(\ref{alg-Snyder-Mx}) and (\ref{alg-Snyder-xp}), respectively. The translation
algebra acts covariantly from the left on the space of commutative functions
as follows. If $\tilde{\phi}(\tilde{x})$ is a noncommutative function and
$\mathbf{1}$ is the identity element of the algebra of commutative functions
over $x_{\mu}$ then%
\begin{equation}
\tilde{\phi}(\tilde{x})\vartriangleright\mathbf{1}=\psi(x), \label{co-alg-def}%
\end{equation}
where $\psi(x)$, in general, differs from $\phi(x)$. Since the noncommutative
functions can be expanded formally in terms of the noncommutative wave
functions $e^{i\left\langle k\tilde{x}\right\rangle }$, the deformed momentum
$K_{\mu}=K_{\mu}(k)$ is defined by the following relation%
\begin{equation}
e^{i\left\langle k\tilde{x}\right\rangle }\vartriangleright\mathbf{1=}%
e^{i\left\langle K\tilde{x}\right\rangle }, \label{k-momentum}%
\end{equation}
with its inverse given by%
\[
e^{i\left\langle K^{-1}(k)\tilde{x}\right\rangle }\vartriangleright
\mathbf{1=}e^{i\left\langle kx\right\rangle }.
\]
The left-action can be extended to products of a finite number of
noncommutative wave functions%
\begin{equation}
e^{i\left\langle K_{1}^{-1}(k_{1})\tilde{x}\right\rangle }e^{i\left\langle
K_{2}^{-1}(k_{2})\tilde{x}\right\rangle }\cdots e^{i\left\langle K_{3}%
^{-1}(k_{3})\tilde{x}\right\rangle }\vartriangleright\mathbf{1}%
=e^{i\left\langle D^{(m)}(k_{m},k_{m-1},\ldots,k_{1})x\right\rangle },
\label{n-k-momentum}%
\end{equation}
where the functions $D^{(m)}(k_{1},k_{2},\ldots,k_{m})$ are defined
recursively as%
\begin{equation}
D_{\mu}^{(m)}(k_{m},k_{m-1},\ldots,k_{1})=D_{\mu}^{(2)}(k_{m},D^{(m-1)}%
(k_{m-1},\ldots,k_{1})). \label{D-functions}%
\end{equation}
In particular, the product of two wave functions determines the $\star$ -
product, the co-product and the anti-pode $S$ of the Poincar\'{e} co-algebra
as follows%
\begin{align}
e^{i\left\langle K_{1}^{-1}(k_{1})\tilde{x}\right\rangle }\star
e^{i\left\langle K_{2}^{-1}(k_{2})\tilde{x}\right\rangle }  &
=e^{i\left\langle D^{(2)}(k_{2},k_{1})x\right\rangle },\label{star-prod}\\
\triangle p_{\mu}  &  =D_{\mu}^{(2)}(p\otimes\mathbf{1},\mathbf{1}\otimes
p),\label{co-prod}\\
D_{\mu}^{(2)}\left(  g,S(g)\right)   &  =0, \label{anti-pode}%
\end{align}
for any element of the deformed Poincar\'{e} group. Thus, the whole structure
of the co-algebra is encoded in the two-functions $D^{(2)}(k_{2},k_{1})$.
These functions depend on the realization of the Snyder geometry. The $\star
$-product and the co-product are nonassociative and noncommutative and that
makes it difficult to construct the field theories. The co-product of the
Lorentz generators takes the following form%
\begin{equation}
\triangle M_{\mu\nu}=M_{\mu\nu}\otimes\mathbf{1}+\mathbf{1}\otimes M_{\mu\nu}.
\label{co-M}%
\end{equation}
The $\star$-product can also be given a representation in terms of
differential operators by taking $p_{\mu}=-i\partial_{\mu}$
\cite{Meljanac:2007xb}. Then one can write%
\begin{equation}
\left(  f\star g\right)  (x)=\lim_{y\rightarrow x}\lim_{z\rightarrow x}%
\exp\left[  i\left\langle \left(  D^{(2)}(p_{y},p_{z})-p_{y}-p_{z}\right)
x\right\rangle \right]  . \label{star-prod-1}%
\end{equation}
The deformed Poincar\'{e} group can be obtained from the co-product of the
translation generators which is compatible with the Lorentz subgroup of the
deformed Poincar\'{e} group according to the relation (\ref{co-M}). In the
realization formalism, any realization represents a deformation of the
Poincar\'{e} algebra that is a generalized Hopf algebra, and that describes
the symmetries of the Snyder geometry with the translation space given by a
deformation of the de Sitter space $SO(1,4)/SO(1,3)$.

\section{Noncommutative fluid in the Snyder space-time}

In this section, we are going to use the realization formalism to derive the
action functional of the noncommutative relativistic fluid in the Snyder
space-time $\mathcal{S}$. The functions over $\mathcal{S}$ can be mapped into
the deformed algebra of the Minkowski space-time $\left(  C^{\infty
}(\mathcal{M}),\star\right)  $. Thus, the action can be represented by a
functional over $\left(  C^{\infty}(\mathcal{M}),\star\right)  $. In the same
way, the deformed Poincar\'{e} group over $\mathcal{S}$ can be mapped
bijectively into the deformed Poincar\'{e} group over $\mathcal{M}$.

\subsection{Action of the noncommutative fluid}

The dynamics of the relativistic ideal fluid in the Minkowski space-time
$\mathcal{M}$ in the Clebsch parametrization can be obtained from an action
functional that depends on the density current and three real fluid potentials
$\phi(x)=\{j^{\mu}(x),\theta(x),\alpha(x),\beta(x)\}$ \cite{Jackiw:2000mm}.
The first step to be taken in order to construct the action of the
noncommutative fluid, is to generalize the potentials to functions
$\tilde{\phi}(\tilde{x})=\{\tilde{j}^{\mu}(x),\tilde{\theta}(\tilde{x}%
),\tilde{\alpha}(\tilde{x}),\tilde{\beta}(\tilde{x})\}$ over the Snyder
space-time that should be identified with the degrees of freedom of the
noncommutative fluid. The correspondence principle in this case is%
\begin{equation}
\lim_{s\rightarrow0}S_{s}[\tilde{\phi}(\tilde{x})]=S[\phi(x)],
\label{cor-princ}%
\end{equation}
where $S_{s}[\tilde{\phi}(\tilde{x})]$ is the action functional of the
noncommutative fluid and $S[\phi(x)]$ is the action of the perfect
relativistic fluid in the Clebsch parametrization. Guided by this principle,
we propose the following Lagrangian for the noncommutative fluid in the Snyder
space-time%
\begin{equation}
\tilde{\mathcal{L}}[\tilde{\theta}(\tilde{x}),\tilde{\alpha}(\tilde{x}%
),\tilde{\beta}(\tilde{x})]=-\tilde{j}^{\mu}(\tilde{x})\left[  \partial_{\mu
}\tilde{\theta}(\tilde{x})+\tilde{\alpha}(\tilde{x})\partial_{\mu}\tilde
{\beta}(\tilde{x})\right]  -\tilde{f}\left(  \sqrt{-\tilde{j}^{\mu}(\tilde
{x})\tilde{j}_{\mu}(\tilde{x})}\right)  , \label{tilde-lagrangian}%
\end{equation}
where $\tilde{j}^{\mu}(\tilde{x})$ is an arbitrary smooth function of the
noncommutative coordinates that generalizes the fluid current and $\tilde{f}$
is an arbitrary smooth function that characterizes the equation of state of a
specific model. According to the realization method discussed in the previous
section, the Lagrangian functional $\tilde{\mathcal{L}}[\tilde{j}^{\mu
}(x),\tilde{\theta}(\tilde{x}),\tilde{\alpha}(\tilde{x}),\tilde{\beta}%
(\tilde{x})]$ is mapped to a functional $\mathcal{L}_{s}[j^{\mu}%
(x),\theta(x),\alpha(x),\beta(x)]$\ that depends on functions from the algebra
$\left(  C^{\infty}(\mathcal{M}),\star\right)  $ where the $\star$-product is
given by the relation (\ref{star-prod-1}). In order to determine the form of
$\mathcal{L}_{s}[j^{\mu}(x),\theta(x),\alpha(x),\beta(x)]$ we perform the
Fourier transformation of the noncommutative potentials%
\begin{equation}
\tilde{\phi}(\tilde{x})=\int[dk]_{s}\hat{\phi}(k)\exp\left(  i\left\langle
K^{-1}(k)\tilde{x}\right\rangle \right)  . \label{Fourier-trans}%
\end{equation}
The integration invariant measure depends on the antipode $S(k_{\mu})=-k_{\mu
}$ which is a realization dependent quantity. However, since the momenta in
different realizations are related by the relations (\ref{momenta}) the
antipode is exactly trivial in all realizations
\cite{Battisti:2010sr,Girelli:2009ii} and the measure takes the following form%
\begin{equation}
\lbrack dk]_{s}=\frac{d^{4}k}{\left(  2\pi\right)  ^{4}}. \label{int-measure}%
\end{equation}
From the Fourier transform (\ref{Fourier-trans}) and the definition of the
$\star$-product (\ref{star-prod-1}) we can derive the first term of the
Lagrangian $\mathcal{L}_{s}[j^{\mu}(x),\theta(x),\alpha(x),\beta(x)]$ as
follows%
\begin{align}
\left(  \tilde{j}^{\mu}(\tilde{x})\partial_{\mu}\tilde{\theta}(\tilde
{x})\right)  \vartriangleright\mathbf{1}  &  =i%
%TCIMACRO{\dint }%
%BeginExpansion
{\displaystyle\int}
%EndExpansion
\frac{d^{4}k_{1}}{\left(  2\pi\right)  ^{4}}\frac{d^{4}k_{2}}{\left(
2\pi\right)  ^{4}}\hat{\jmath}^{\mu}(k_{1})k_{2,\mu}\hat{\theta}(k_{2})\left(
\exp\left(  i\left\langle K^{-1}(k_{1})\tilde{x}\right\rangle \right)
\exp\left(  i\left\langle k_{2}\tilde{x}\right\rangle \right)  \right)
\nonumber\\
&  =i%
%TCIMACRO{\dint }%
%BeginExpansion
{\displaystyle\int}
%EndExpansion
\frac{d^{4}k_{1}}{\left(  2\pi\right)  ^{4}}\frac{d^{4}k_{2}}{\left(
2\pi\right)  ^{4}}\hat{\jmath}^{\mu}(k_{1})k_{2,\mu}\hat{\theta}(k_{2}%
)\exp\left(  i\left\langle D^{(2)}(k_{1},k_{2})x\right\rangle \right)
\label{Four-trans-j-th}\\
&  =j^{\mu}(x)\star\partial_{\mu}\theta(x).\nonumber
\end{align}
The last product is defined on the algebra $\left(  C^{\infty}(\mathcal{M}%
),\star\right)  $. The second term of the Lagrangian $\tilde{\mathcal{L}%
}[\tilde{\theta}(\tilde{x}),\tilde{\alpha}(\tilde{x}),\tilde{\beta}(\tilde
{x})]$ can be mapped into $\left(  C^{\infty}(\mathcal{M}),\star\right)  $ in
exactly the same way. The third term involves a triple $\star$-product. Its
Fourier transform can be calculated by using the relation (\ref{D-functions}).
After some algebraic manipulations, the following result is obtained%
\begin{align}
\left(  \tilde{j}^{\mu}(\tilde{x})\tilde{\alpha}(\tilde{x})\partial_{\mu
}\tilde{\beta}(\tilde{x})\right)  \vartriangleright\mathbf{1}  &  =i%
%TCIMACRO{\dint }%
%BeginExpansion
{\displaystyle\int}
%EndExpansion
\left(  \prod\limits_{m=1}^{3}\frac{d^{4}k_{m}}{\left(  2\pi\right)  ^{4}%
}\right)  \hat{\jmath}^{\mu}(k_{1})\hat{\alpha}(k_{2})k_{3,\mu}\hat{\beta
}(k_{3})\exp\left(  i\left\langle D^{(3)}(k_{3},k_{2},k_{1})x\right\rangle
\right) \nonumber\\
&  =j^{\mu}(x)\star\left(  \alpha(x)\star\partial_{\mu}\beta(x)\right)  .
\label{Four-trans-j-a-b}%
\end{align}
The relations (\ref{tilde-lagrangian}), (\ref{Four-trans-j-th}) and
(\ref{Four-trans-j-a-b}) lead to the following action of the noncommutative
fluid defined on the algebra $\left(  C^{\infty}(\mathcal{M}),\star\right)  $%
\begin{align}
S_{s}\left[  j^{\mu}(x),\theta(x),\alpha(x),\beta(x)\right]   &  =%
%TCIMACRO{\dint }%
%BeginExpansion
{\displaystyle\int}
%EndExpansion
d^{4}x\tilde{\mathcal{L}}[\tilde{\theta}(\tilde{x}),\tilde{\alpha}(\tilde
{x}),\tilde{\beta}(\tilde{x})]\vartriangleright\mathbf{1}\nonumber\\
&  =%
%TCIMACRO{\dint }%
%BeginExpansion
{\displaystyle\int}
%EndExpansion
d^{4}x\left[  -j^{\mu}(x)\star\left[  \partial_{\mu}\theta(x)+\alpha
(x)\star\partial_{\mu}\beta(x)\right]  - f_{s}\left(  \sqrt{-j^{\mu}(x) \star
j_{\mu}(x)}\right)  \right]  , \label{action-star}%
\end{align}
where in the second term from (\ref{action-star}) the $\star$-product from the
square bracket should be computed firstly. The relationship between $\tilde
{f}$ and $f_{s}$ is given by the map (\ref{co-alg-def}). Since the function
$\tilde{f}$ is arbitrary, the action (\ref{action-star}) describes a class of
noncommutative fluids parametrized by $\alpha$, $\beta$ and $f_{s}$ for any
given value of $s$ \footnote{This is different from the noncommutative fluid
in the K\"{a}hler parametrization in which the action describes a class of
fluids parametrized by $f_{\lambda}$ \emph{and} the K\"{a}hler potential
$K_{\lambda}$ where $\lambda$ is the noncommutative parameter
\cite{Holender:2011px}.}. The equation (\ref{action-star}) can be used to
construct the noncommutative deformation of a commutative fluid model
characterized by a particular function $f$ by deforming $f$ to $f_{s}$ such
that $\lim_{s\rightarrow0}f_{s}=f$. Then it is a simple exercise to verify
that the action (\ref{action-star}) satisfies the corresponding principle
(\ref{cor-princ}).

\subsection{Deformed Poincar\'{e} transformations}

The common point of view adopted to define the physical quantities associated
to a noncommutative field theory is that they should be associated to the
group of transformations of the noncommutative structure underlying the
theory. According to this point of view, the infinitesimal variation of the
action $\delta_{\varepsilon}S_{s}$ under the deformed Poincar\'{e} algebra
should define physical quantities relevant to the noncommutative fluid
described by (\ref{action-star}). Note that, in general, the variation
$\delta_{\varepsilon}$ viewed as an operator on $\left(  C^{\infty
}(\mathcal{M}),\star\right)  $ is linear but does not necessarily satisfy the
Leibniz's condition. The variation of the action under the deformed
Poincar\'{e} transformations is defined by the usual relation%
\begin{equation}
\delta_{\varepsilon}S_{s}=S_{s}(\varepsilon)-S_{s}, \label{var-act-def}%
\end{equation}
where $S_{s}(\varepsilon)$ represents the action with all variables acted upon
by the infinitesimal deformed Poincar\'{e} transformations%
\begin{equation}
x_{\mu}\rightarrow x_{\mu}+\delta_{\varepsilon}x_{\mu}, \label{gen-transf}%
\end{equation}
where $\delta_{\varepsilon}x$ is defined by (\ref{alg-Snyder-Mx}) and
(\ref{alg-Snyder-xp}). In the first case, the parameter $\varepsilon_{\mu}$ is
an infinitesimal constant vector on $\mathcal{M}$ while in the second case it
is an infinitesimal antisymmetric constant matrix $\varepsilon_{\mu\nu
}=-\varepsilon_{\nu\mu}$. The transformation (\ref{gen-transf}) induces a map
between the translated functions from $\mathcal{F}(\mathcal{S})$ and $\left(
C^{\infty}(\mathcal{M}),\star\right)  $%
\begin{equation}
\tilde{\phi}(\tilde{x}+\delta_{\varepsilon}\tilde{x})\vartriangleright
\mathbf{1}=\psi(x+\delta_{\varepsilon}x), \label{var-map}%
\end{equation}
where $\psi$ is the function defined by the equation (\ref{co-alg-def}).

Let us consider the deformed translations%
\begin{equation}
\delta_{\varepsilon}\tilde{x}_{\mu}=\left[  \tilde{x}_{\mu},\left\langle
\varepsilon p\right\rangle \right]  . \label{deformed-tran}%
\end{equation}
For any realization, the variation (\ref{deformed-tran}) induces a
transformation in $\mathcal{M}$ which, in general, is not the commutative
translation. This can be seen by inverting the relation (\ref{rep-1}) and by
calculating the variation of $x_{\mu}$ from%
\begin{equation}
\delta_{\varepsilon}x_{\mu}=\left[  x_{\mu},\left\langle \varepsilon
p\right\rangle \right]  =i\varepsilon_{\mu}+isp_{\mu}\left\langle \varepsilon
p\right\rangle \frac{\varphi_{2}(A)}{\varphi_{1}(A)}. \label{mod-translation}%
\end{equation}
As can be easily checked, the transformation (\ref{mod-translation}) leaves
the volume element invariant. After some algebra, we can show that the
relation (\ref{var-act-def}) takes the following form%
\begin{equation}
\delta_{\varepsilon}S_{s}=%
%TCIMACRO{\dint }%
%BeginExpansion
{\displaystyle\int}
%EndExpansion
d^{4}x\left[  \mathcal{L}_{s}\left(  x+\delta_{\varepsilon}x\right)
-\mathcal{L}_{s}\left(  x\right)  \right]  , \label{var-act-1}%
\end{equation}
where $\mathcal{L}_{s}\left(  x+\delta_{\varepsilon}x\right)  $ is the
Lagrangian from (\ref{action-star}) with the $\star$-product computed at
$x+\delta_{\varepsilon}x$.

The deformed rotations have the following form%
\begin{equation}
\delta_{\varepsilon}\tilde{x}_{\mu}=\left[  \tilde{x}_{\mu},\left\langle
\varepsilon M\right\rangle \right]  , \label{deformed-rot}%
\end{equation}
where $\left\langle \varepsilon M\right\rangle =\varepsilon^{\mu\nu}M_{\mu\nu
}$. The induced transformation in $\mathcal{M}$ can be obtained from the
inverse of the relation (\ref{rep-1}) and takes the following form%
\begin{equation}
\delta_{\varepsilon}x_{\mu}=\left[  x_{\mu},\left\langle \varepsilon
M\right\rangle \right]  =-\frac{2}{\varphi_{1}(A)}\varepsilon_{\mu}^{\nu
}\tilde{x}_{\nu}. \label{mod-rot}%
\end{equation}
The volume element is invariant under the transformation (\ref{gen-transf})
with (\ref{mod-rot}). Therefore, the variation of the action under
(\ref{deformed-rot}) has the form (\ref{var-act-1}).

In general, there is no simple expression that describes the variation
$\delta_{\varepsilon}S_{s}$ due to the nonassociativity of the $\star
$-product. Indeed, one can compute the $\star$-product at $x+\delta
_{\varepsilon}x$ and compare it with the Baker-Campbell-Hausdorff formula. We
consider the mapping of the additive subgroup by the exponential at
right\footnote{The additive subgroup is mapped to right and left product of
exponentials which are not isomorphic to each other due to the
noncommutativity of the product.}. Then after some calculations one can show
that under the deformed translations%
\begin{align}
&  \exp\left[  i\left\langle \left(  D^{(2)}(p_{y},p_{z})-p_{y}-p_{z}\right)
x\right\rangle \right]  \exp\left[  i\left\langle \left(  D^{(2)}(p_{y}%
,p_{z})-p_{y}-p_{z}\right)  \delta_{\varepsilon}x\right\rangle \right]
=\nonumber\\
&  \exp\left\{  i\left\langle \left(  D^{(2)}(p_{y},p_{z})-p_{y}-p_{z}\right)
\left(  x+\delta_{\varepsilon}x\right)  \right\rangle \right. \nonumber\\
&  -\frac{s\varphi_{2}(A)}{2\varphi_{1}(A)}\left(  \eta_{\mu\nu}\left\langle
\varepsilon p_{x}\right\rangle +p_{\nu}^{x}\varepsilon_{\mu}\right)  \left(
D^{(2)\mu}(p_{y},p_{z})-p_{y}^{\mu}-p_{z}^{\mu}\right)  \left(  D^{(2)\nu
}(p_{y},p_{z})-p_{y}^{\nu}-p_{z}^{\nu}\right) \nonumber\\
&  \left.  -\frac{is\varphi_{2}(A)}{12\varphi_{1}(A)}\left(  \varepsilon_{\mu
}\eta_{\rho\sigma}+\eta_{\mu\nu}\varepsilon_{\rho}\right)  \left(  D^{(2)\rho
}(p_{y},p_{z})-p_{y}^{\rho}-p_{z}^{\rho}\right)  \left(  D^{(2)\nu}%
(p_{y},p_{z})-p_{y}^{\nu}-p_{z}^{\nu}\right)  \left(  D^{(2)\mu}(p_{y}%
,p_{z})-p_{y}^{\mu}-p_{z}^{\mu}\right)  \right\}  . \label{BCH-formula}%
\end{align}
As can be seen, the exponential that defines the $\star$-product at
$x+\delta_{\varepsilon}x$ does factorize in the Maggiore realization but not
in the Snyder's. In general, even if the exponential factorizes, the $\star
$-product does not. This behavior is not restricted to the relativistic
fluid. Actually, it is the result of the structure of the noncommutative
algebra given by the relations (\ref{alg-Snyder-x})-(\ref{alg-Snyder-Mp})\ and
it is expected to hold for any field theory. Similar conclusions can be drawn
for the deformed rotations. Apparently, the difficulties generated by the
nonassociativity of the $\star$ - product could be circumvented by defining
the variation of the action functional as generated by the operator
$\delta_{\varepsilon}$ instead of (\ref{var-act-def}) with the following action%
\begin{equation}
\delta_{\varepsilon}S_{s}=[%
%TCIMACRO{\dint }%
%BeginExpansion
{\displaystyle\int}
%EndExpansion
d^{4}x\tilde{\mathcal{L}}[\tilde{j}^{\mu}(x),\tilde{\theta}(\tilde{x}%
),\tilde{\alpha}(\tilde{x}),\tilde{\beta}(\tilde{x})]\vartriangleright
\mathbf{1},\left\langle \varepsilon G\right\rangle ], \label{op-var}%
\end{equation}
where $G$ is either $p_{\mu}$ or $M_{\mu\nu}$. However, the problems related
to the nonassociativity return in the form of the variation of the $\star
$-products from $\mathcal{L}_{s}[j^{\mu}(x),\theta(x),\alpha(x),\beta(x)]$ as
can be verified easily.

A very important consequence of the nonassociativity of the $\star$-product is
that it makes it difficult to calculate and even to define relevant physical
quantities associated to the fields, such as the energy and the momentum.
Nevertheless, some important properties of the quantities associated with the
variations $\delta_{\varepsilon}x_{\mu}$ can be derived in the general case.
To this end, we write the deformed Poincar\'{e} transformations
(\ref{mod-translation}) and (\ref{mod-rot}) as%
\begin{align}
\delta_{\nu}x_{\mu} &  =i\left(  \eta_{\nu\mu}+sp_{\nu}p_{\mu}\frac
{\varphi_{2}(A)}{\varphi_{1}(A)}\right)  ,\label{var-tr}\\
\delta_{\rho\sigma}x_{\mu} &  =\left[  \eta_{\rho\sigma}\left(  x_{\sigma
}+s\left\langle xp\right\rangle p_{\sigma}\right)  \frac{\varphi_{2}%
(A)}{\varphi_{1}(A)}\right]  -\left[  \eta_{\sigma\mu}\left(  x_{\rho
}+s\left\langle xp\right\rangle p_{\rho}\right)  \frac{\varphi_{2}(A)}%
{\varphi_{1}(A)}\right]  .\label{var-rot}%
\end{align}
Then one can show by direct calculations that the variation of the action
under the deformed transformations $\delta_{\varepsilon}x_{\mu}$ produces the
following equation%
\begin{equation}
\partial_{\mu}\left(  \Theta^{\mu\nu}(\phi)\delta_{\varepsilon}x_{\nu}\right)
=\partial^{\mu}\left(  \mathcal{L}\delta_{\varepsilon}x_{\mu}\right)
,\label{A-1}%
\end{equation}
where $\Theta^{\mu\nu}(\phi)$ is a functional of the fluid potentials and
their derivatives up to the third order. Since the (\ref{var-rot}) depends
linearly on $x_{\mu}$, it follows that the equations of motion alone are not
sufficient to guarantee the conservation of the quantities described by the
functions $\Theta^{\mu\nu}(\phi)$ associated to $\delta_{\rho\sigma}x_{\mu}$.
On the other hand, the right hand side of the equation (\ref{var-tr}) is
independent of $x_{\mu}$. One can check that the following quantity associated
to the deformed translations%
\begin{equation}
T_{\nu}^{\mu}=\Theta_{\sigma}^{\mu}(\phi)-\mathcal{L}\eta_{\nu}^{\mu
}\label{A-2}%
\end{equation}
is conserved. Note that, in general, the functions $\Theta^{\mu\nu}(\phi)$
that correspond to the translation are different from the ones derived from
the rotations. $T_{\nu}^{\mu}$ represents the variation of the action $S_{s}$
under the deformed translation. Therefore, it can be interpreted as being the
energy-momentum tensor of the noncommutative fluid in an arbitrary
realization. Note that the tensor $T_{\mu\nu}=\eta_{\mu\rho}T_{\nu}^{\rho}$
does not have a definite symmetry. Actually, a symmetric energy-momentum
tensor can be obtained by coupling the fluid with a $c$ - number metric
$g_{\mu\nu}$ and by deriving the action with respect to it (see
\cite{Holender:2011px}). However, by this procedure information about the
noncommutative properties of the fluid could be lost due to the contraction
between the antisymmetric components of the $\star$-product and the metric.
The invariance of the theory under the noncommutative translations has been
discussed and used in the literature to define the energy-momentum tensor of
different field theories
\cite{Grimstrup:2002xs,Bellucci:2003ud,Ghosh:2003ka,Bertolami:2003nm,Banerjee:2003vc,Bellucci:2004ura,Agostini:2006nc,Mariz:2006kp,Ghosh:2007ai,Agostini:2007ia}%
. The equation (\ref{mod-translation}) generalizes the deformed translations
to the realization formalism which treats simultaneously various
noncommutative spaces as we have seen in the previous section. From this point
of view, the equation (\ref{A-2}) represent a generalization of the previous
results within the realization formalism.

\section{Conclusions and Discussions}

In this paper, we have constructed for the first time a model of the
noncommutative fluid on the Snyder space-time. To this end, we have used the
realization formalism of the noncommutative spaces and we have generalized the
action functional formulation of the relativistic perfect fluid in the
Minkowski space-time. This model is important to understanding the behavior
of the effective (or long wave) degrees of freedom on noncommutative spaces.
It provides a new class of noncommutative field theories with deformed
Poincar\'{e} symmetry that generalizes the field theories on the commutative
space-time in the first order formulation. By using the realization maps from
the algebra $\mathcal{F}(\mathcal{S})$ to the $\left(  C^{\infty}%
(\mathcal{M}),\star\right)  $, we have obtained a representation of a large
class of noncommutative fluids parametrized by three arbitrary functions
$\alpha(x)$, $\beta(x)$ and $f(x)$ in terms of the deformed algebra of smooth
functions on the Minkowski space-time. In this formulation, the fluid dynamics
is given by the equation of motion of the fluid potentials viewed as fields on
the Snyder space and subjected to the conservation of the energy-momentum
tensor. Establishing these equations in the general case is a difficult task
due to the interactions among the fluid potentials that involve infinitely
many derivatives of fields.\ This particular structure is the result of the
action of the $\star$-product which is neither commutative nor associative.
However, it is possible to study the theory perturbatively in the
noncommutative parameter $s$. Nevertheless, even at the first order, the
equations of motion are highly non-linear even in the simplest case of the
model that reduces to the irrotational fluid in the commutative limit. It is
an important and interesting problem to study these equation and to determine
their integrability and possible solutions. We hope to report on these topics elsewhere.

\noindent\textbf{Acknowledgments} The work of M. C. B. A. was partially
supported by the CNPq Grant: 306276/2009-7. M. A. S. and I. V. V. acknowledge
financial support from IFT-UNESP. I.\ V. V. would like to thank to N.
Berkovits for hospitality at ICTP-SAIFR where this work was accomplished and
to H. Nastase and A. Mikhailov for useful discussions.

\end{document}